\documentclass[acmsmall, review=false, nonacm=true, printfolios=true]{acmart}

\AtBeginDocument{%
  \providecommand\BibTeX{{%
    \normalfont B\kern-0.5em{\scshape i\kern-0.25em b}\kern-0.8em\TeX}}}

\setcopyright{acmcopyright}
\copyrightyear{2018}
\acmYear{2018}
\acmDOI{10.1145/1122445.1122456}

\acmConference[Woodstock '18]{Woodstock '18: ACM Symposium on Neural
  Gaze Detection}{June 03--05, 2018}{Woodstock, NY}
\acmBooktitle{Woodstock '18: ACM Symposium on Neural Gaze Detection,
  June 03--05, 2018, Woodstock, NY}
\acmPrice{15.00}
\acmISBN{978-1-4503-XXXX-X/18/06}

\usepackage[ruled,vlined,linesnumbered]{algorithm2e}

\SetKwComment{tcp}{}{}
\newcommand{\ioformat}[1]{{\hfill \texttt{#1}}}
\SetKwInput{KwInput}{\parbox{1.5cm}{Input}}
\SetKwInput{KwOutput}{\parbox{1.5cm}{Output}}

\usepackage{booktabs}
\usepackage{lipsum}
\usepackage{tikz}
\usepackage{adjustbox}   
\usepackage{subcaption}  
\usepackage{layouts}  
\usepackage{multirow}

\def\decompname{CP}
\def\CC{{C\nolinebreak[4]\hspace{-.05em}\raisebox{.2ex}{\tiny\bf ++}}}

\newcommand{\mathcalbf}[1]{\pmb{\mathcal{#1}}}

\newcommand{\vect}[1]{\ensuremath{\mathbf{#1}}}

\newcommand{\mat}[1]{\ensuremath{\mathbf{#1}}}
\newcommand{\matt}[2]{\ensuremath{\mat{#1}_{#2}}}
\newcommand{\set}[1]{\ensuremath{\mathcal{#1}}}
\newcommand{\tensor}[1]{\ensuremath{\mathcalbf{#1}}}
\newcommand{\unfolding}[2]{\ensuremath{\mat{#1}_{(#2)}}}

\newcommand{\inr}[1]{\ensuremath{\in \mathbb{R}^{#1}}}

\newcommand{\factor}[2]{\ensuremath{\mathbf{U}_{#2}^{(#1)}}}
\newcommand{\krp}[2]{\ensuremath{\mathbf{M}_{#2}^{(#1)}}}
\newcommand{\hadam}[2]{\ensuremath{\mathbf{H}_{#2}^{(#1)}}}

\newcommand{\multifactor}[2]{\ensuremath{\mathbf{\overline{U}}_{#2}^{(#1)}}}
\newcommand{\multikrp}[2]{\ensuremath{\mathbf{\overline{M}}_{#2}^{(#1)}}}

\newcommand{\range}[2]{#1, $\ldots$ , #2}

\begin{document}

\title{Accelerating jackknife resampling for the Canonical Polyadic Decomposition}
\renewcommand{\shorttitle}{Accelerating jackknife for CPD}

\author{Christos Psarras}
\orcid{0000-0001-6057-7491}
\affiliation{%
  \institution{RWTH Aachen University}
  \streetaddress{Schinkelstr. 2}
  \city{Aachen}
  \state{North Rhine-Westphalia}
  \postcode{52062}
  \country{Germany}}
\email{christos.psarras@rwth-aachen.de}

\author{Lars Karlsson}
\affiliation{%
  \institution{Ume{\aa} Universitet}
  \streetaddress{Plan 4, MIT-huset}
  \city{Ume{\aa}}
  \country{Sweden}}
\email{larsk@cs.umu.se}

\author{Rasmus Bro}
\orcid{0000-0002-7641-4854}
\affiliation{%
    \institution{University of Copenhagen}
    \streetaddress{Rolighedsvej 26}
    \city{Copenhagen}
    \state{}
    \postcode{1958 Frederiksberg C}
    \country{Denmark}}
\email{rb@food.ku.dk}

\author{Paolo Bientinesi}
\affiliation{%
	\institution{Ume{\aa} Universitet}
	\streetaddress{Plan 4, MIT-huset}
	\city{Ume{\aa}}
	\country{Sweden}}
\email{larsk@cs.umu.se}

\renewcommand{\shortauthors}{Psarras, Karlsson, Bro and Bientinesi}

\begin{abstract}
 The Canonical Polyadic (CP) tensor decomposition is frequently used as a model in applications in a variety of different fields.
 Using jackknife resampling to estimate parameter uncertainties is often desirable but results in an increase of the already high computational cost.
 Upon observation that the resampled tensors, though different, are nearly identical,
 we show that it is possible to extend the recently proposed Concurrent ALS (CALS) technique to a jackknife resampling scenario.
 This extension gives access to the computational efficiency advantage of CALS for the price of a modest increase (typically a few percent) in the number of floating point operations.
 Numerical experiments on both synthetic and real-world datasets demonstrate that the new workflow based on a CALS extension
 can be several times faster than a straightforward workflow where the jackknife submodels are processed individually.
\end{abstract}

\ccsdesc[500]{Mathematics of computing~Mathematical software performance}
\ccsdesc[500]{Software and its engineering~Software performance}

\keywords{jackknife, Tensors, Decomposition, CP, ALS}

\maketitle


\section{Introduction}
\label{sec:01-Introduction}

The CP model is used increasingly across a large diversity of fields.
One of the fields in which CP is commonly applied is chemistry~\cite{Murphy:2013, wiberg2004}, where there is often a
need for estimating not only the parameters of the model, but also the associated uncertainty of those parameters~\cite{farrance2012}.
In fact, in some areas it is a dogma  that an estimate without an uncertainty is not a result.
A common approach for estimating uncertainties of the parameters of CP models is through resampling, such as bootstrapping or jackknifing~\cite{riu:2003, kiers:2004}.
The latter has added benefits, e.g., for variable selection~\cite{martens2000} and outlier detection\cite{riu:2003}. 
Here we consider a new technique, JK-CALS, that increases the performance of jackknife resampling applied to CP by more efficiently utilizing the computer's memory hierarchy.

The basic concept of jackknife is somewhat related to cross-validation.
Let $\tensor{T}\inr{I_1 \times \cdots \times I_N}$ be a tensor, and \range{\matt{U}{1}}{\matt{U}{N}} the factor matrices of a CP model.
Let us also make the assumption (typical in many applications) that the first mode corresponds to independent samples, and
all the other modes correspond to variables.
For the most basic type of jackknifing, namely Leave-One-Out (LOO)\footnote{Henceforth, when we mention jackknifing we
  imply LOO jackknifing, unless otherwise stated.}, one sample (out of $I_1$) is left out 
at a time (resulting in a tensor with only $I_1 - 1$ samples)
and a model is fitted to the remaining data; we refer to that model as a \emph{submodel}.
All samples are left out exactly once, resulting in $I_{1}$ distinct submodels.
Each submodel provides an estimate of the parameters of the overall model.
For example, each submodel provides an estimate of the factor (or loading) matrix $\matt{U}{2}$.
From these $I_{1}$ estimates it is possible to calculate the variance (or bias) of the overall loading matrix (the one
obtained from all samples).
One complication comes from some indeterminacies with CP that need to be taken into account.
For example, when one (or more) samples are removed from the initial tensor, the order of components in the submodel may change;
this phenomenon is explained and a solution is proposed in~\cite{riu:2003}.

Recently, the authors proposed a technique, Concurrent ALS (CALS)~\cite{psarras2020}, that can fit \emph{multiple} CP
models to the \emph{same} underlying tensor more rapidly than regular ALS.
CALS achieves better performance not by altering the numerics but by utilizing the computer's memory hierarchy more efficiently than regular ALS.
However, the CALS technique cannot be directly applied to jackknife resampling, since the $I_1$
submodels are fitted to \emph{different} tensors. 
In this paper, we extend the idea that underpins CALS to jackknife resampling.
The new technique takes advantage of the fact that the $I_1$ resampled tensors are \emph{nearly} identical.
At the price of a modest increase in arithmetic operations, the technique allows for more efficient fitting of the CP submodels and thus improved overall performance of a jackknife workflow.
In applications in which the number of components in the CP model is relatively low, the technique can significantly
reduce the overall time to solution.\\

\paragraph*{Contributions}
\begin{itemize}
\item An efficient technique, JK-CALS, for performing jackknife resampling of CP models.
  The technique is based on an extension of CALS to \emph{nearly} identical tensors.
  To the best of our knowledge, this is the first attempt at accelerating jackknife resampling of CP models. 

\item Numerical experiments demonstrate that JK-CALS can lead to performance gains in a jackknife resampling workflow. 

\item Theoretical analysis shows that the technique generalizes from leave-one-out to delete-$d$ jackknife with a modest (less than a factor of two) increase in arithmetic.

\item A \CC{} library with support for GPU acceleration and a Matlab interface.\\
\end{itemize}

\paragraph*{Organization}
The rest of the paper is organized as follows.
In Section~\ref{sec:02-related-work}, we provide an overview of related research.
In Section~\ref{sec:03-cp-als-cals-and-jackknife}, we review the standard \decompname{}-ALS and CALS algorithms, as well as jackknife applied to CP.
We describe the technique which enables us to use CALS to compute jackknife more efficiently in Section~\ref{sec:04-accelerating-jackknife-using-cals}.
In Section~\ref{sec:05-experiments} we demonstrate the efficiency of our proposed technique, by applying it to perform jackknife resampling to CP models that have been fitted to artificial and real tensors.
Finally, in Section~\ref{sec:06-conclusion}, we conclude the paper and provide insights for further research.



\section{Related Work}
\label{sec:02-related-work}

Two popular techniques for uncertainty estimation for CP models are bootstrap and jackknife~\cite{westad:2015, kiers:2004, riu:2003}.
The main difference is that jackknife resamples \emph{without} replacement whereas bootstrap resamples \emph{with} replacement.
Bootstrap frequently involves more submodels than jackknife and is therefore more expensive.
The term jackknife typically refers to leave-one-out jackknife, where only one observation is removed when resampling. More than one observation can be removed at a time~\cite{peddada:1993}; a variation commonly called delete-$d$ jackknife.
When applied to CP, jackknife has certain benefits over bootstrap, e.g., for variable selection~\cite{martens2000} and outlier detection\cite{riu:2003}.

Jackknife requires fitting multiple submodels.
A straightforward way of accelerating jackknife is to separately accelerate the fitting of each submodel, e.g., using a faster implementation.
The simplest and most extensively used numerical method for fitting CP models is the Alternating Least Squares (CP-ALS) method.
Alternative methods for fitting CP models include eigendecomposition-based methods~\cite{Sanchez:1986} and
gradient-based (all-at-once) optimization methods~\cite{2011_acar}.

Several techniques have been proposed to accelerate CP-ALS.
Line search~\cite{2008_rajih} and extrapolation~\cite{2019_ang} aim to reduce the number of iterations until convergence.
Randomization-based techniques have also been proposed.
These target very large tensors, and either randomly sample the tensor~\cite{2016_vervliet} or the Khatri-Rao product~\cite{2018_battaglino}, to reduce their size and, by extension, the overall amount of computation.
Similarly, compression-based techniques replace the target tensor with a compressed version, thus also reducing the amount of computation during fitting~\cite{1998c_bro}.
The CP model of the reduced tensor is inflated to correspond to a model of the original tensor.

Several projects offer high-performance implementations of CP-ALS, for example, Cyclops~\cite{Solomonik:2013},
PLANC~\cite{Ramakrishnan:2016}, Partensor~\cite{Lourakis:2018}, SPLATT~\cite{Smith:2015}, and Genten~\cite{Phipps:2019}.
For a more comprehensive list of software implementing some variant of CP-ALS, refer to~\cite{psarras:2021}.

Similar to the present work, there have been attempts at accelerating jackknife although (to the best of our knowledge) not in the context of CP.
In~\cite{buzas:1997}, the high computational cost of jackknife is tackled by using a numerical approximation that
requires fewer operations at the price of lower accuracy.
In~\cite{belotti:2020}, a general-purpose  routine for fast jackknife estimation is presented.
Some estimators (often linear ones) have leave-one-out
formulas that allow for fast computation of the estimator after leaving one sample out.
Jackknife is thus accelerated by computing the estimator on the full set and then systematically applying the leave-one-out formula.
In~\cite{hinkle:1996}, a similar technique is studied.
Jackknife computes an estimator on $s$ distinct subsets of the $s$ samples.
Any two of these subsets differ by only one sample, i.e., any one subset can be obtained from any other by replacing one and only one element.
Some estimators have a fast updating formula, which can rapidly transform an estimator for one subset to an estimator for another subset. 
Jackknife is thus accelerated by computing the estimator from scratch on the first subset and then repeatedly updating the estimator using this fast updating formula.



\section{CP-ALS, CALS and jackknife}
\label{sec:03-cp-als-cals-and-jackknife}

In this section, we first specify the notation to be used throughout the paper, 
we then review the \decompname{}-ALS and CALS techniques, 
and finally we describe jackknife resampling applied to CP.\\ 

\subsection{Notation} \label{sec:03:notation}
For vectors and matrices, we use bold lowercase and uppercase roman letters, respectively, e.g., \vect{v} and \mat{U}.
For tensors, we follow the notation in~\cite{Kolda:2009};
specifically, we use bold calligraphic fonts, e.g., \tensor{T}.
The order (number of indices or modes) of a tensor is denoted by uppercase roman letters, e.g., $N$.
For each mode $n \in \{  1, 2, \ldots, N \}$, a tensor \tensor{T} can be unfolded (matricized) into a matrix, denoted by \unfolding{T}{n}, where the columns are the mode-$n$ fibers of \tensor{T}, i.e., the vectors obtained by fixing all indices except for mode~$n$.
Sets are denoted by calligraphic fonts, e.g., $\mathcal S$.
Given two matrices $\mat A$ and $\mat B$ with the same number of columns, the Khatri-Rao product, denoted by $\mat A
\odot \mat B$, is the column-wise Kronecker product of $\mat A$ and $\mat B$. Finally, the unary operator $\oplus$, when 
applied to a matrix, denotes the scalar which is the sum of all matrix elements.
\\

\subsection{CP-ALS} \label{sec:03:als}
The standard alternating least-squares method for \decompname{} is shown in Algorithm~\ref{alg:als} (CP-ALS).
The input consists of a target tensor \tensor{T}.
The output consists of a CP model represented by a sequence of factor matrices \range{\matt{U}{1}}{\matt{U}{N}}.
The algorithm repeatedly updates the factor matrices one by one in sequence until either of the following criteria are met:
a) the fit of the model to the target tensor falls below a certain tolerance threshold, or b) a maximum number of iterations has been reached.
To update a specific factor matrix \matt{U}{n}, the gradient of the least-squares objective function with respect to that factor matrix is set to zero and the resulting linear least-squares problem is solved directly from the normal equations. 
This entails computing the Matricized Tensor Times Khatri-Rao Product (MTTKRP) (line~\ref{alg:als:mttkrp}), which is the product between the mode-$n$ unfolding \unfolding{T}{n} and the Khatri-Rao Product (KRP) of all factor matrices except \matt{U}{n}.
The MTTKRP is followed by the Hadamard product of the Gramians of each factor matrix (${\matt{U}{i}}^{T}\matt{U}{i}$) in line~\ref{alg:als:hadamard}.
Factor matrix \matt{U}{n} is updated by solving the linear system in line~\ref{alg:als:update}.
At the completion of an iteration, i.e., a full pass over all $N$ modes, the error between the model and the target tensor
is computed (line~\ref{alg:als:error}) using the efficient formula derived in \cite{Phan:2013}.

\begin{algorithm}
    \SetAlgoLined
    \DontPrintSemicolon
    \KwInput{%
      \parbox{30mm}{\tensor{T}\inr{I_1 \times \cdots \times I_N}}
      \ioformat{\small The target tensor}
    }
    \KwOutput{%
      \parbox{30mm}{$\matt{U}{1}, \ldots, \matt{U}{N}$} \ioformat{\small The fitted factor matrices}}
    Initialize the factor matrices $\matt{U}{1}, \ldots, \matt{U}{N}$\;
    \Repeat{convergence detected or maximum number of iterations reached}
    {
        \For{\range{$n = 1, 2$}{$N$}}{
            $\matt{M}{n} \leftarrow \matt{T}{(n)} (\odot_{i\neq n} \matt{U}{i})$ \tcp*[r]{\small MTTKRP} \label{alg:als:mttkrp}
            $\matt{H}{n} \leftarrow \ast_{i\neq n}({\matt{U}{i}}^{T}\matt{U}{i})$ \tcp*[r]{\small Hadamard product of Gramians} \label{alg:als:hadamard}
            $\matt{U}{n} \leftarrow \matt{M}{n} {\matt{H}{n}}^{\dagger}$ \tcp*[r]{\small ${\matt{H}{n}}^{\dagger}$: pseudoinverse of \matt{H}{n}} \label{alg:als:update}
        }
        $e \leftarrow ||\tensor{T}||^{2} - (\oplus (\matt{H}{N} \ast({\matt{U}{N}}^{T}\matt{U}{N}))) - 2 (\oplus (\matt{U}{N}
        \ast \matt{M}{N}))$ \tcp*[r]{\small Error calculation} \label{alg:als:error}
    }
    \caption{\decompname{}-ALS: Alternating least squares method for \decompname{} decomposition.}
    \label{alg:als}
\end{algorithm}

Assuming a small number of components ($R$), the most expensive step is the MTTKRP (line~\ref{alg:als:mttkrp}).
This step involves $2 R \prod_{i} I_{i}$ FLOPs (ignoring, for the sake of simplicity, the lower order of FLOPs required for the computation of the KRP).
The operation touches slightly more than $\prod_{i} I_{i}$ memory locations, resulting in an arithmetic intensity less than $2 R$ FLOPs per memory reference.
Thus, unless $R$ is sufficiently large, the speed of the computation will be limited by the memory bandwidth rather than the speed of the processor.
The CP-ALS algorithm is inherently memory-bound for small $R$, regardless of how it is implemented. 

\begin{figure}[h]
    \centering
    \input{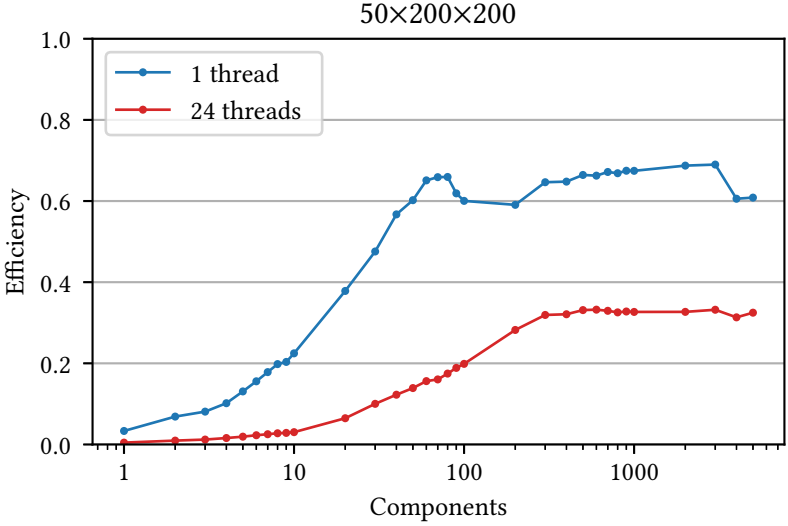}
    \caption{Efficiency of MTTKRP on a $50 \times 200 \times 200$ tensor for an increasing number of components.
    Note that in the multi-threaded execution, the theoretical peak performance increases while the total number of operations to be performed
    stays the same (as in the single-threaded case); this explains the drop in efficiency per thread.}
    \label{fig:MTTKRP_rank_sweep}
\end{figure}

The impact on performance of the memory-bound nature of MTTKRP is demonstrated in Fig.~\ref{fig:MTTKRP_rank_sweep}, which shows the computational efficiency of a particular implementation of MTTKRP as a function of the number of components (for a tensor of size $50 \times 200 \times 200$).
Efficiency is defined as the ratio of the performance achieved by MTTKRP (in FLOPs/sec), relative to the Theoretical Peak
Performance (TPP, see below) of the machine, i.e., 
\begin{displaymath}
\textsc{efficiency} = \frac{\textsc{performance}}{\textsc{tpp}} = \frac{\textsc{\#flops} / \textsc{time}}{\textsc{tpp}}.
\end{displaymath}
The TPP of a machine is defined as the maximum number of (double precision) floating point operations the machine can perform in one second.
Table~\ref{tab:TPP} shows the TPP for our particular machine (see Sec.~\ref{sec:05-experiments} for details).
\begin{table}[h]
    \centering
    \begin{tabular}{lccc}
    	\toprule
    	System             & TPP (GFlops/sec) & threads & frequency per core (Ghz) \\ \midrule
    	\multirow{2}*{CPU} &       112        &    1    &           3.5            \\
    	                   &       1536       &   24    &            2             \\ \bottomrule
    \end{tabular}
    \caption{Theoretical Peak Performance (TPP) for a particular machine. Due to the decrease in the peak frequency per core when all 24 cores are used, the TPP for 24 cores is less than $24 \times$ the TPP for 1 core.}
    \label{tab:TPP}
\end{table}
In Fig.~\ref{fig:MTTKRP_rank_sweep},
we see that the efficiency of MTTKRP tends to increase with the number of components, $R$, until eventually reaching a plateau.
On this machine, the plateau is $R \geq 60$ at $\approx 70\%$ efficiency for one thread and $R \geq 300$ at $\approx 35\%$ efficiency for 24 threads. 
For $R \leq 20$, which is common in applications, the efficiency is well below the TPP.\\

\subsection{Concurrent ALS (CALS)} \label{sec:03:cals}

When fitting \emph{multiple} CP models to the \emph{same} underlying tensor, the Concurrent ALS (CALS) technique can improve the efficiency if the number of components is not large enough for CP-ALS to reach its performance plateau~\cite{psarras2020}.
A need to fit multiple models to the same tensor arises, for example, when trying different initial guesses or when trying different numbers of components.

The gist of CALS can be summarized as follows (see~\cite{psarras2020} for details).
Suppose  $K$ independent instances of CP-ALS have to be executed on the same underlying tensor.
Rather than running them sequentially or in parallel, run them in lock-step fashion as follows.
Advance every CP-ALS process one iteration before proceeding to the next iteration.
One CALS iteration entails $K$ CP-ALS iterations (one iteration per model).
Each CP-ALS iteration in turn contains one MTTKRP operation, so one CALS iteration also entails $K$ MTTKRP operations.
But these MTTKRPs all involve the same tensor and can therefore be fused into one bigger MTTKRP operation (see Eq.~3 of~\cite{psarras2020}).
The performance of the fused MTTKRP depends on the sum total of components, i.e., $\sum_{i=1}^{K} R_{i}$, where $R_{i}$ is the number of components in model~$i$.
Due to the performance profile of MTTKRP (see Fig.~\ref{fig:MTTKRP_rank_sweep}), the fused MTTKRP is expected to be more efficient than each of the $K$ smaller operations it replaces.

The following example illustrates the impact on efficiency of MTTKRP fusion.
Given a target tensor of size $50 \times 200 \times 200$, $K=50$ models to fit, and $R_i = 5$ components in each model, the efficiency for
each of the $K$ MTTKRPs in CP-ALS is about $15\%$ ($3\%$) for 1 (24) threads (see Fig.\ref{fig:MTTKRP_rank_sweep}).
The efficiency of the fused MTTKRP in CALS will be as observed for $R = \sum_{i=1}^{K} R_{i} = 250$, i.e.,  $60\%$ ($30\%$) for 1 (24) threads.
Since the MTTKRP operation dominates the cost, CALS is expected to be $\approx 4 \times$ ($\approx 10 \times$) faster than CP-ALS for 1 (24) threads.\\

\subsection{Jackknife} \label{sec:03:jackknife}

Algorithm~\ref{alg:jackknife} shows a baseline (inefficient) application of leave-one-out jackknife resampling to a CP model.
For details, see~\cite{riu:2003}.
The inputs are a target tensor \tensor{T}, an overall CP model $P$ fitted to all of \tensor{T}, and
a sampled mode $\hat{n} \in \{ 1, 2, \ldots, N \}$.
For each sample $p \in \{ 1, 2, \ldots, I_{\hat{n}} \}$, the algorihm removes the slice corresponding to the sample from tensor \tensor{T} (line~\ref{alg:jk:tensor_subsample}) and model $P$ (line~\ref{alg:jk:model_subsample}) and fits a reduced model $P_{-p}$ (lines~\ref{alg:jk:model_subsample}--\ref{alg:jk:perm_scale}) to the reduced tensor \tensor{\hat{T}} using regular CP-ALS. 
After fitting all submodels, the standard deviation of every factor matrix except \matt{U}{\hat{n}} is computed from the $I_{\hat{n}}$ submodels in $\set{Q}$ (line~\ref{alg:jk:std}).
The only costly part of Algorithm~\ref{alg:jackknife} is the repeated calls to CP-ALS in line~\ref{alg:jk:fitting}.

\begin{algorithm}
    \SetAlgoLined
    \DontPrintSemicolon
    \KwInput{%
        \parbox{35mm}{\tensor{T}\inr{I_1 \times \cdots \times I_N}} \ioformat{\small The target tensor} \newline%
        \parbox{35mm}{$P = \matt{U}{1}, \ldots, \matt{U}{N}$} \ioformat{\small A CP model fitted to \tensor{T}} \newline%
        \parbox{35mm}{$\hat{n}$} \ioformat{\small The sampled mode}
    }
    \KwOutput{%
        \parbox{35mm}{$\matt{S}{1}, \ldots, \matt{S}{N}$}
        \ioformat{\small Uncertainty of each element of each factor matrix of P}
    }
    
    $\set{Q} \leftarrow \emptyset$ \tcp*[]{\small Set containing fitted jackknife models}
    \For(\tcp*[f]{\small For every index $p$ in mode $\hat{n}$}){
        $p \in \{ 1, 2, \ldots, I_{\hat{n}} \}$}
    {
        $\tensor{T}_{-p} \leftarrow $ remove the slice with index $p$ in mode $\hat{n}$ from tensor \tensor{T}\label{alg:jk:tensor_subsample}\;
        $P_{-p} \leftarrow $ remove row $p$ from factor matrix \matt{U}{\hat{n}} of $P$\label{alg:jk:model_subsample}\;
        $\hat{P}_{-p} \leftarrow $ \texttt{cp\_als}($\tensor{T}_{-p}$, $P_{-p}$)\label{alg:jk:fitting}\;
        $\hat{P}_{-p} \leftarrow $ permutation and scale adjustment of $\hat{P}_{-p}$\label{alg:jk:perm_scale}\;
        $\set{Q} \leftarrow \set{Q} \cup \{ \hat{P}_{-p} \}$ \;
    }
    \For(\tcp*[f]{\small For every mode $n$ except $\hat{n}$}){
        $n \in \{ 1, 2, \ldots, N \} \setminus \{ \hat{n} \}$}
    {
        $\matt{S}{n} \leftarrow$ standard deviation of factor matrix \matt{U}{n} in $\set{Q}$\label{alg:jk:std}\;
    }
    \caption{JK-ALS: An algorithm that performs (LOO) jackknife resampling on a CP model.}
    \label{alg:jackknife}
\end{algorithm}



\section{Accelerating jackknife by using CALS}
\label{sec:04-accelerating-jackknife-using-cals}

The straightforward application of jackknife to CP in Algorithm~\ref{alg:jackknife} involves $I_{\hat{n}}$ independent calls to CP-ALS on \emph{nearly} the same tensor.
Since the tensors are not exactly the same, CALS~\cite{psarras2020} cannot be directly applied.
In this section, we show how one can rewrite Algorithm~\ref{alg:jackknife} in such a way that CALS \emph{can} be applied.
There is an associated overhead due to extra computation, but we will show that the overhead is modest (less than a 100\% increase and typically only a few percent increase).\\

\subsection{JK-CALS: Jackknife extension of CALS} \label{sec:04:proof}

Let \tensor{T} be an $N$-mode tensor with a corresponding CP model $\matt{A}{1}, \ldots, \matt{A}{N}$.
Let $\mathcalbf{\hat{T}}$ be identical to $\mathcalbf{T}$ except for one sample (with index $p$) removed from the sampled mode $\hat{n} \in \{ 1, 2, \ldots, N \}$.
Let $\matt{\hat{A}}{1}, \ldots, \matt{\hat{A}}{N}$ be the CP submodel corresponding to the resampled tensor \tensor{T}.

When fitting a CP model to $\mathcalbf{T}$ using CP-ALS, the MTTKRP for mode $n$ is given by
\begin{equation}
\mathbf{M}_n \gets \unfolding{T}{n} (\matt{A}{N} \odot \dots \odot \matt{A}{n+1} \odot \matt{A}{n-1} \odot \dots \odot \matt{A}{1}).
\label{eqn:MTTKRP}
\end{equation}
Similarly, when fitting a model to $\mathcalbf{\hat{T}}$, the MTTKRP for mode $n$ is given by
\begin{equation}
\mathbf{\hat{M}}_{n} \gets \unfolding{\hat{T}}{n} (\matt{\hat{A}}{N} \odot \dots \odot \matt{\hat{A}}{n+1} \odot \matt{\hat{A}}{n-1} \odot \dots \odot \matt{\hat{A}}{1}).
\label{eqn:MTTKRP_jack}
\end{equation}

Can $\mathbf{\hat{M}}_{n}$ be computed from $\unfolding{T}{n}$ instead of $\unfolding{\hat{T}}{n}$?
As we will see, the answer is yes. 
We separate two cases: $n = \hat{n}$ and $n \neq \hat{n}$.

{\bfseries Case I: $n = \hat{n}$.}
The slice of $\mathcalbf{T}$ removed when resampling corresponds to a \emph{row} of the unfolding $\unfolding{T}{n} = \unfolding{T}{\hat{n}}$.
To see this, note that element $\tensor{T}(i_1, i_2, \dots, i_N)$ corresponds to element $\unfolding{T}{n}(i_n, j)$ of its
mode-$n$ unfolding~\cite{Kolda:2009}, where
\begin{equation} 
j = 1 + \sum_{\substack{k=1 \\ k\neq n}}^{N} (i_k - 1) \prod_{\substack{m=1 \\ m \neq n}}^{k-1} I_m.
\label{eqn:kolda_mapping}
\end{equation}
When we remove sample $p$, then $\unfolding{\hat{T}}{n}$ will be identical to $\unfolding{T}{n}$ except that row $p$ from the latter is missing in the former.
In other words, $\unfolding{\hat{T}}{n} = \mathbf{E}_{p} \unfolding{T}{n}$, where $\mathbf{E}_{p}$ is the matrix that removes row $p$. 
We can therefore compute $\matt{\hat{M}}{n}$ by replacing $\unfolding{\hat{T}}{n}$ with $\unfolding{T}{n}$ in (\ref{eqn:MTTKRP_jack}) and then discarding row $p$ from the result:
\begin{displaymath}
  \mathbf{\hat{M}}_{n} \gets \mathbf{E}_{p} ( \unfolding{T}{n} (\matt{\hat{A}}{N} \odot \dots \odot \matt{\hat{A}}{n+1} \odot \matt{\hat{A}}{n-1} \odot \dots \odot \matt{\hat{A}}{1}) ).
\end{displaymath}

{\bfseries Case II: $n \neq \hat{n}$.}
The slice of $\mathcalbf{T}$ removed when resampling corresponds to a \emph{set of columns} in the unfolding $\unfolding{T}{n}$.
One could in principle remove these columns to obtain $\unfolding{\hat{T}}{n}$. 
But instead of explicitly removing sample $p$ from \tensor{T}, we can simply zero out the corresponding slice of \tensor{T}.
To give the CP model matching dimensions, we need only insert a row of zeros at index $p$ in factor matrix $\hat{n}$.
Crucially, the zeroing out of slice $p$ is superfluous.
In the MTTKRP, the elements that should have been zeroed out will be multiplied with zeros in the Khatri-Rao product generated by the row of zeros insert in factor matrix $\hat{n}$. 
Thus, to compute $\mathbf{\hat{M}}_{n}$ in (\ref{eqn:MTTKRP_jack}) we (a) replace $\unfolding{\hat{T}}{n}$ with $\unfolding{T}{n}$ and (b) insert a row of zeros at index $p$ in factor matrix $\matt{\hat{A}}{\hat{n}}$.

In summary, we have shown that it is possible to compute $\mathbf{\hat{M}}_{n}$ in (\ref{eqn:MTTKRP_jack})
without referencing the reduced tensor.
There is an overhead associate with extra arithmetic. 
For the case $n = \hat{n}$, we compute numbers that are later discarded.
For the case $n \neq \hat{n}$, we do some arithmetic with zeros. 

Based on the observations above, the CALS algorithm~\cite{psarras2020} can be modified to facilitate the concurrent fitting of all jackknife submodels.
Algorithm~\ref{alg:cals_jackknife} incorporates the necessary changes.
In the end, extending CALS to support jackknife comes down to these localized changes (colored red in Algorithm~\ref{alg:cals_jackknife}):
\begin{itemize}
\item Insert a row of zeros in one of the factor matrices,
\item periodically zero out the padded row to keep it zero, and
\item adjust the error formula to compute the submodel error. 
\end{itemize}

We remark that JK-CALS can be straightforwardly extended to delete-$d$ jackknife.
Instead of padding and periodically zeroing out one row, we pad and periodically zero out $d$ rows. \\

\begin{algorithm}
    \SetAlgoLined
    \DontPrintSemicolon
    \KwInput{\tensor{T}\inr{I_1 \times \cdots \times I_N} \ioformat{\small The target tensor}\newline%
        $\hat{n}$ \ioformat{\small The sampled mode}}
    \KwOutput{$\matt{U}{1}^{(p)}, \ldots, \matt{U}{N}^{(p)}$ for $p = 1, 2, \ldots, I_{\hat{n}}$ \ioformat{\small The fitted submodels}}
    Initialize the submodels\;
    \For(\tcp*[f]{\small Initialize one factor multi-matrix for each mode}){\range{$n = 1, 2$}{$N$}}{\label{alg:start-jk-1}
        \For{\range{$p = 1, 2$}{$I_{\hat{n}}$}}{
            {\color{red}\eIf{$n = \hat{n}$}{$\multifactor{n}{|p} \leftarrow \matt{U}{n}^{(p)}$ with a row of zeros inserted at index $p$ \label{alg:cals_jk:multifactor0}}
                {$\multifactor{n}{|p} \leftarrow \matt{U}{n}^{(p)}$ \label{alg:cals_jk:multifactor1}}}
        }
    }\label{alg:stop-jk-1}
    \Repeat(\tcp*[f]{\small Concurrently run $I_{\hat{n}}$ instances of \decompname{}-ALS\label{alg:jk-loop-modes}}){convergence detected for all instances or maximum number of iterations reached}{
        \For{\range{$n = 1, 2$}{$N$}}{
            $\multikrp{n}{} \leftarrow \mat{T}_{(n)}(\odot_{i\neq n} \multifactor{i}{})$ \label{alg:cals_jk:mttkrp}\;
            \For{\range{$p = 1, 2$}{$I_{\hat{n}}$}}{
                $\hadam{n}{p} \leftarrow \ast_{i\neq n}({\multifactor{i}{|p}}^{T}\multifactor{i}{|p}) \label{alg:cals_jk:hadamard}$\;
                $\multifactor{n}{|p} \leftarrow \multikrp{n}{|p}{\hadam{n}{p}}^{\dagger}$ \label{alg:cals_jk:update}\;
                {\color{red}\If{$n = \hat{n}$}{$\multifactor{n}{|p} \leftarrow$ zero out row $p$ of \multifactor{n}{|p}} \label{alg:cals_jk:multifactor}}
            }
        }
        \For{\range{$p = 1, 2$}{$I_{\hat{n}}$}}{
            $e \leftarrow {\color{red}||\tensor{T}_{-p}||^{2}} - (\oplus (\hadam{N}{p}\ast({\factor{N}{|p}}^{T}\factor{N}{|p})))
            - 2 (\oplus (\factor{N}{|p} \ast \krp{N}{|p}))$ \tcp*[r]{\small Error calculation} \label{alg:cals_jk:error}
        }
    }
    \caption{JK-CALS: Concurrent alternating least squares method for jackknife estimation.}
    \label{alg:cals_jackknife}
\end{algorithm}

\subsection{Performance considerations}

While Algorithm~\ref{alg:cals_jackknife} benefits from improved MTTKRP efficiency, the padding results in extra arithmetic operations.
Let $d$ denote the number of removed samples ($d = 1$ corresponds to leave-one-out).
For the sake of simplicity, assume that
the integer $d$ divides $I_{\hat{n}}$.
There are $I_{\hat{n}} / d$ submodels, each with $R$ components.
The only costly part is the MTTKRP.

The MTTKRPs in JK-ALS (for all submodels combined) requires
\begin{displaymath}
  \left(\frac{I_{\hat{n}}}{d} \right)
  \left( 2 R (I_{\hat{n}} - d) \prod_{\substack{i  = 1 \\ i\neq \hat{n}}}^{N} I_{i} \right)  
\end{displaymath}
FLOPs.
Meanwhile, the fused MTTKRP in JK-CALS requires
\begin{displaymath}
  2 \left( \frac{I_{\hat{n}}}{d} R \right) \prod_{i=1}^{N} I_{i}
\end{displaymath}
FLOPs.
The ratio of the latter to the former comes down to
\begin{displaymath}
  \frac{I_{\hat{n}}}{I_{\hat{n}} - d} \leq 2,
\end{displaymath}
since $d \leq I_{\hat{n}} / 2$ in delete-$d$ jackknife.
Thus, in the worst case, JK-CALS requires less than twice the FLOPs of JK-ALS.
More typically, the overhead is negligible.



\section{Experiments}
\label{sec:05-experiments}

We investigate the performance benefits of the JK-CALS algorithm to perform jackknife resampling on a CP model through two sets of experiments.
In the first set of experiments, we focus on the scalability of the algorithm, with respect to both problem size and number of processor cores.
For this purpose, we use synthetic datasets of increasing volume, mimicking the shape of real datasets.
In the second set of experiments, we illustrate JK-CALS's practical impact by using it to perform jackknife
resampling on two tensors arising in actual applications.

All experiments were conducted using a Linux-based system with an Intel\textregistered{} Xeon\textregistered{} Platinum 8160 Processor (Turbo Boost enabled, Hyper-Threading disabled), which contains 24 physical cores split in 2 NUMA regions of 12 cores each.
The system also contains an Nvidia Tesla V100 GPU\footnote{Driver version: 470.57.02, CUDA Version: 11.2}.
The experiments were conducted with double precision arithmetic and we report results for 1 thread, 24 threads (two NUMA
regions), and the GPU (with 24 CPU threads).
The source code (available online\footnote{\href{https://github.com/HPAC/CP-CALS/tree/jackknife}{https://github.com/HPAC/CP-CALS/tree/jackknife}}) was compiled using GCC\footnote{GCC version 9} and linked to the Intel\textregistered{} Math Kernel Library\footnote{MKL version 19.0}.\\

\subsection{Scalability analysis}
In this first experiment, we use three synthetic tensors of size $50 \times m \times m$ with $m \in \{100, 200, 400\}$, referred
to as ``small'', ``medium'' and ``large'' tensors, respectively.
The samples are in the first mode. 
The other modes contain variables.
The number of samples is kept low, since leave-one-out jackknife is usually performed on a small number of samples (usually $< 100$), while there can be arbitrarily many variables.

For each tensor, we perform jackknife on four models with varying number of components ($R \in \{3, 5, 7, 9\}$).
This range of component counts is typical in applications. In practice, it is often the case that multiple models are fitted to the target tensor, and many of those models are then further analyzed using jackknife.
For this reason, we perform jackknife on each model individually, as well as to all models simultaneously
(denoted by ``All'' in the figures), to better simulate multiple real-world application scenarios.
In this experiment, the termination criteria based on maximum number of iterations and tolerance are ignored;
instead, all models are forced to go through exactly 100 iterations, typically a small number of iterations
for small values of tolerance (i.e., most models require more than 100 iterations).
The reason for this choice is that we aim to isolate the performance difference of the methods tested; therefore, we maintain a consistent amount of workload throughout the experiment.
(Tolerance and maximum number of iterations are instead used later on in the application experiments.)

For comparison, we perform jackknife using three methods: JK-ALS, JK-OALS and JK-CALS.
JK-OALS uses OpenMP to take advantage of the inherent parallelism when fitting multiple submodels. This method is only used for multi-threaded and GPU experiments, and we
are only going to focus on its performance, ignoring the memory overhead associated with it. 

\begin{figure}
    \centering
    \adjustbox{max width=\textwidth}{
    \input{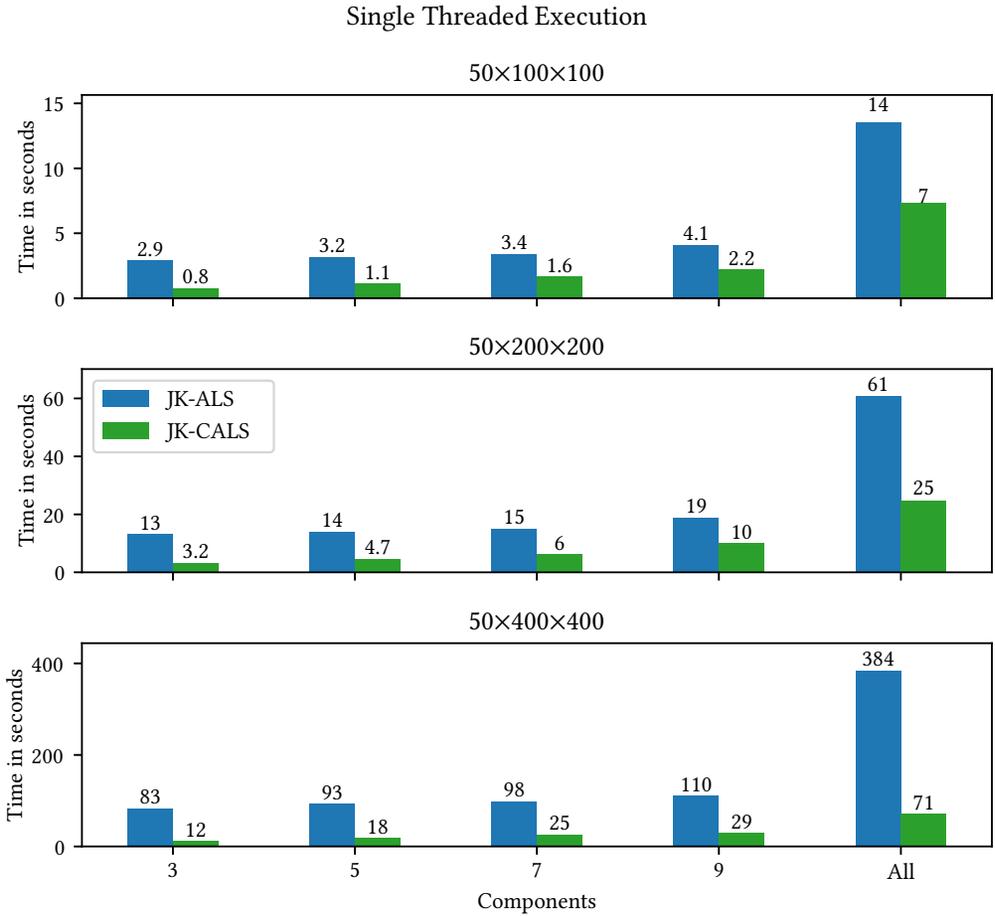}}
    \caption{
      Execution time for single-threaded jackknife resampling applied to three different tensors (small,
      medium, and large in the top, middle, and bottom panels, respectively), and different number of components
      (from left to right: $R \in \{ 3, 5, 7, 9 \}$, and ``All'', which represents doing jackknife to all four models simultaneously).
    }
    \label{fig:artif_1}
\end{figure}

Fig.~\ref{fig:artif_1} shows results for single threaded execution; in this case, JK-OALS is absent.
JK-CALS consistently outperforms JK-ALS for all tensor sizes and workloads.
Specifically, for any fixed amount of workload---i.e., a model of a specific number of components---JK-CALS exhibits increasing speedups compared to JK-ALS, as the tensor size increases.
For example, for a model with 5 components, JK-CALS is $2.9$, $3$, $5.2$ times faster than JK-ALS, for the small, medium and large tensor sizes, respectively.

\begin{figure}
    \centering
    \adjustbox{max width=\textwidth}{
    \input{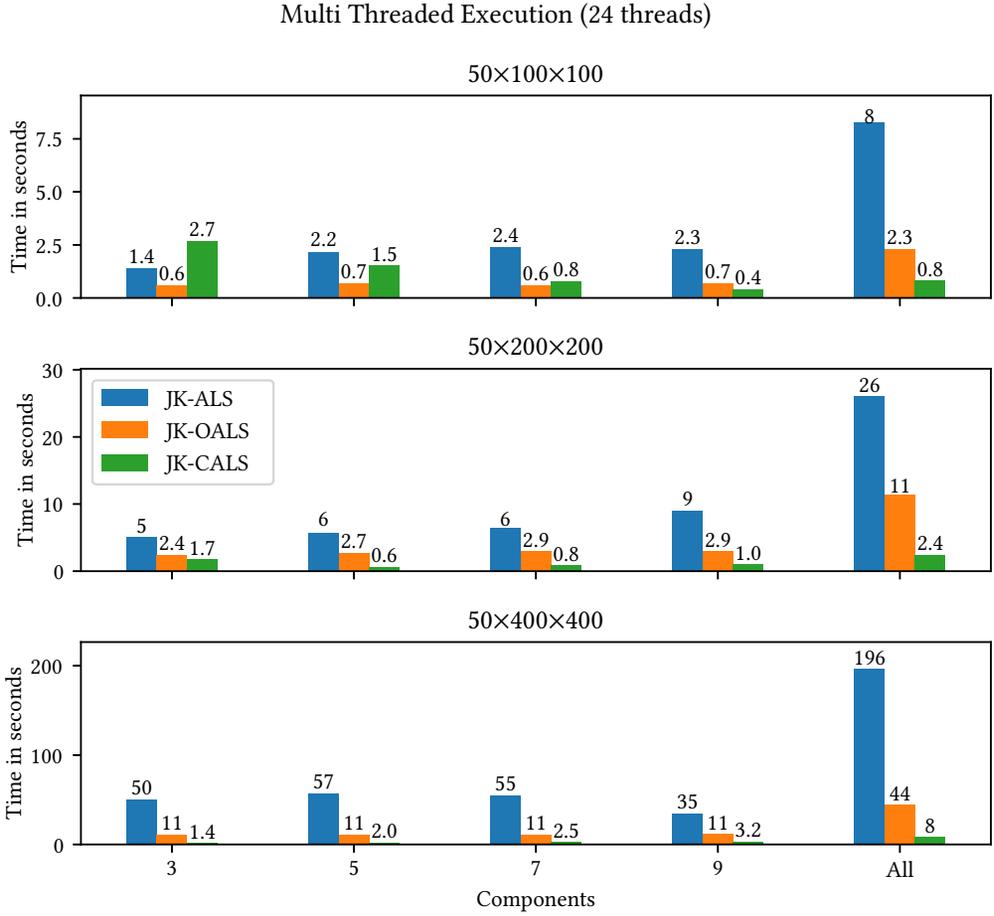}}
    \caption{
      Execution time for multi-threaded (24 threads) jackknife resampling applied to three different tensors (small,
      medium, and large in the top, middle, and bottom panels, respectively), and different number of components
      (from left to right: $R \in \{ 3, 5, 7, 9 \}$, and ``All'', which represents doing jackknife to all four models simultaneously).
    }
    \label{fig:artif_24}
\end{figure}

Fig.~\ref{fig:artif_24} shows results for multi-threaded execution, using 24 threads.
In this case, JK-CALS outperforms the other two implementations (JK-ALS and JK-OALS) for the medium and large tensors,
for all workloads (number of components), 
exhibiting speedups up to $35\times$ and $8\times$ compared to JK-ALS and JK-OALS, respectively.
For the small tensor ($50 \times 100 \times 100$) and small workloads ($R\leq7$), JK-CALS is outperformed by JK-OALS; for $R=3$, it is also outperformed by JK-ALS.
Investigating this further, for the small tensor and $R=3$ and $5$, the parallel speedup (the ratio between single threaded and multi-threaded execution time) of JK-CALS is $0.3\times$ and $0.7\times$ for 24 threads.
However, for $12$ threads, the corresponding timings are $0.28$ and $0.27$ seconds, resulting in speedups of $2.7\times$
and $3.7\times$ respectively. 
This points to two main reasons for the observed performance of JK-CALS in these cases:
a) the amount of available computational resources (24 threads) is disproportionately high compared to the
volume of computation to be performed
and b) because of the small amount of overall computation, the small overhead associated with the CALS methodology
becomes more significant.

That being said, even for the small tensor, as the amount of workload increases---already for a single model with 9
components---JK-CALS again becomes the fastest method. 
Finally, similarly to the single threaded case, as the size of the tensor increases, so do the speedups achieved by JK-CALS over the other two methods.

\begin{figure}
    \centering
    \adjustbox{max width=\textwidth}{
    \input{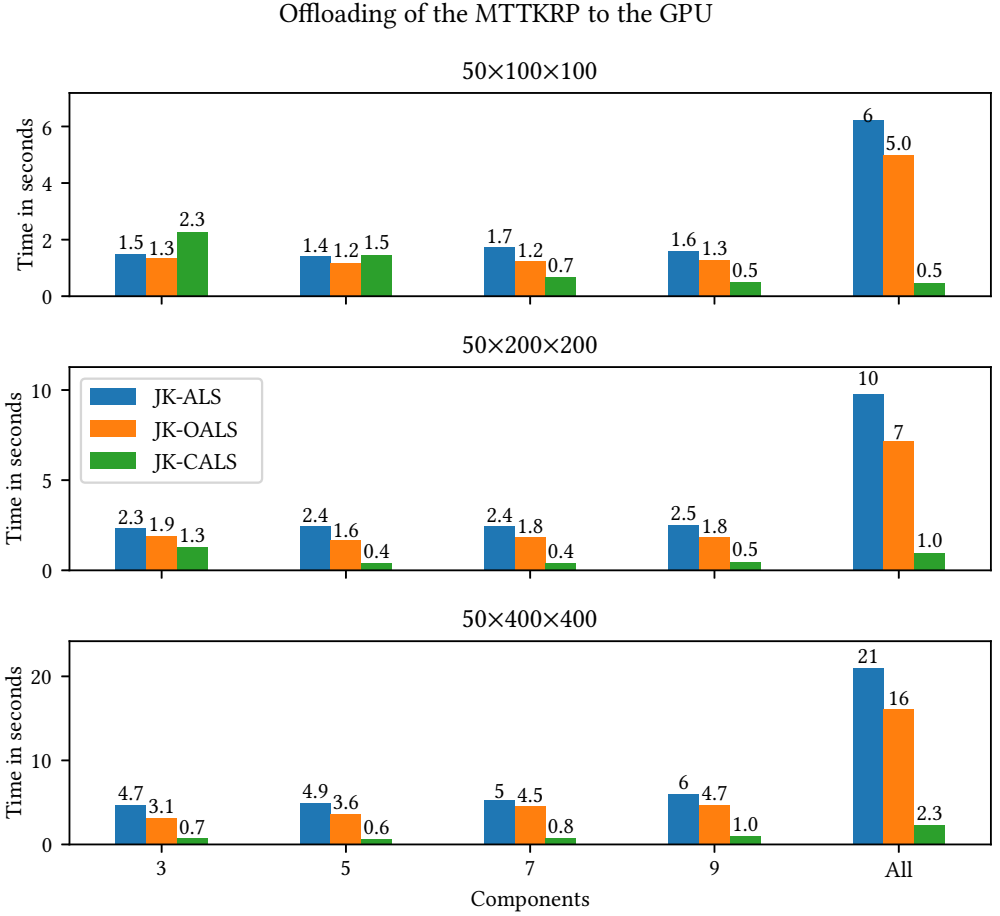}}
    \caption{
      Execution time for GPU + multi-threaded 
      (GPU + 24 threads) jackknife resampling applied to three different tensors (small,
      medium, and large in the top, middle, and bottom panels, respectively), and different number of components
      (from left to right: $R \in \{ 3, 5, 7, 9 \}$, and ``All'', which represents doing jackknife to all four models simultaneously).
    }
    \label{fig:artif_GPU}
\end{figure}

Fig.~\ref{fig:artif_GPU} shows results when the GPU is used to perform MTTKRP for all three methods; in this case, all 24 threads are used on the CPU.
For the small tensor and small workloads ($R \leq 5$), there is not enough computation to warrant the shipping of data to and from the GPU, resulting in higher execution times compared to multi-threaded execution;
for all other cases, all methods have reduced execution time when using the GPU compared to the execution on 24 threads.
Furthermore, in those cases, JK-CALS is consistently faster than its counterparts, exhibiting the largest speedups when
the workload is at its highest (``All''), with values of $10\times$, $7\times$, $7\times$ compared to JK-OALS, and $12\times$, $10\times$, $9\times$ compared to JK-ALS, for the small, medium and large tensors, respectively.\\

\subsection{Real-world applications}

\begin{figure}
    \centering
    \adjustbox{max width=\textwidth}{
    \input{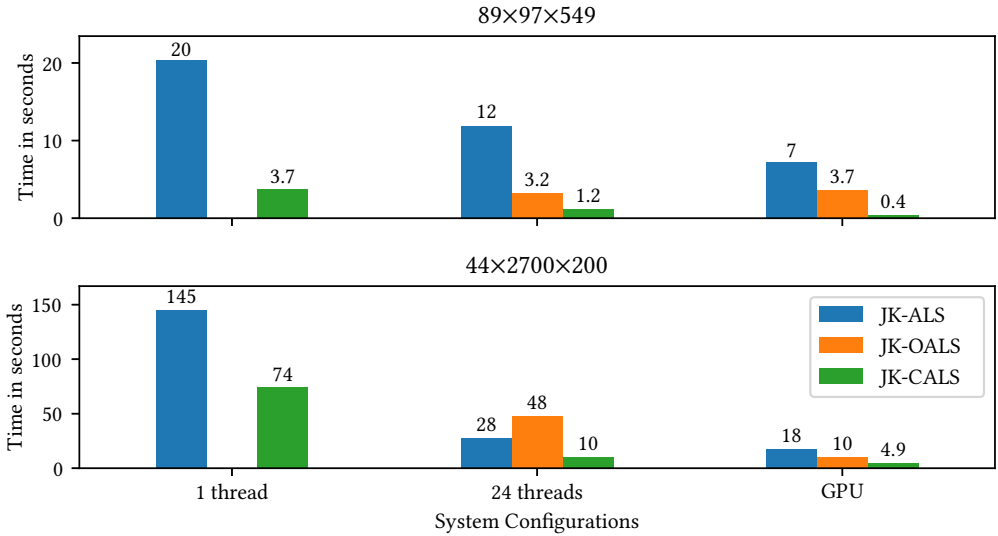}}
    \caption{
      Execution time for jackknife resampling applied to two applications tensors.
      For tensor $89 \times 97 \times 549$, whose expected rank is 5, three models with $R \in \{ 4, 5, 6 \}$ were fitted and
      then jackknife was applied to them (i.e., the ``All'' group from the previous section).
      Similarly, for tensor $44 \times 2700 \times 200$, whose expected rank is 20, three models with $R \in \{ 19, 20, 21 \}$
      were fitted and then jackknifed. 
      In both cases, tolerance and maximum number of iterations were set to $10^{-6}$ and $1000$ respectively.
    }
    \label{fig:real}
\end{figure}

In this second experiment, we
selected two tensors of size $89 \times 97 \times 549$ and $44 \times 2700 \times 200$ from the field of Chemometrics~\cite{acar:2008,Skov:2008}.
In this field it is common to fit multiple, randomly initialized models in a range of low components (e.g. $R \in \{ 1, 2, \ldots, 20 \}$, $10$--$20$ models for each $R$, and then analyze (e.g., using jackknife)
those models that might be of particular interest (often those with components close to the expected rank of the target tensor);
in the tensors we consider, the expected rank is $5$ and $20$, respectively.
To mimic the typical workflow of practitioners, we fitted three models to each tensor, of components $R \in \{ 4, 5, 6 \}$ and
$R \in \{ 19, 20, 21 \}$, respectively, and used the three methods (JK-ALS, JK-OALS and JK-CALS) to apply jackknife resampling to the fitted models.
The values for tolerance and maximum number of iterations were set according to typical values for the particular field, namely $10^{-6}$ and $1000$, respectively.

In Fig.~\ref{fig:real} we report the execution time for 1 thread, 24 threads, and GPU + 24 threads.
For both datasets and for all configurations,
JK-CALS is faster than the other two methods. 
Specifically, when compared to JK-ALS over the two tensors, JK-CALS achieves speedups of $5.4\times$ and $2\times$ for
single threaded execution, $10\times$ and $2.8\times$ for 24-threaded execution.
Similarly, when compared to JK-OALS, JK-CALS achieves speedups of $2.7\times$ and $4.8\times$ for
24-threaded execution.
Finally, JK-CALS takes advantage of the GPU the most,
exhibiting speedups of $17.5\times$ and $3.7\times$ over JK-ALS, and $9\times$ and $2\times$ over JK-OALS, for GPU execution.



\section{Conclusion}
\label{sec:06-conclusion}

Jackknife resampling of CP models is useful for estimating uncertainties, but the computation requires fitting multiple submodels and is therefore computationally expensive.
We presented a new technique for implementing jackknife that better utilizes the computer's memory hierarchy. 
The technique is based on a novel extension of the Concurrent ALS (CALS) algorithm for fitting multiple CP models to the same underlying tensor, first introduced in \cite{psarras2020}.
The new technique has a modest arithmetic overhead that is bounded above by factor of two in the worst case. 
Numerical experiments on both synthetic and real-world datasets using a multicore processor paired with a GPU demonstrated that the proposed algorithm can be several times faster than a straightforward implementation of jackknife resampling based on multiple calls to a regular CP-ALS implementation.

Future work includes extending the software to support delete-$d$ jackknife. 


\section*{Conflict of Interest Statement}
The authors declare that the research was conducted in the absence of any commercial or financial relationships that could be construed as a potential conflict of interest.

\section*{Author Contributions}
CP drafted the main manuscript text, developed the source code, performed the experiments and prepared all figures.
LK and PB revised the main manuscript text.
CP, LK, and PB discussed and formulated the jackknife extension of CALS.
CP, LK, RB, and PB discussed and formulated the experiments.
LK, RB, and CP discussed the related work section.
PB oversaw the entire process.
All authors reviewed and approved the final version of the manuscript.

\section*{Data Availability Statement}
The source code and datasets used in this study are available online: \href{https://github.com/HPAC/CP-CALS/tree/jackknife}{https://github.com/HPAC/CP-CALS/tree/jackknife}.

\begin{acks}
This work was funded by the Deutsche Forschungsgemeinschaft (DFG, German Research Foundation) – 333849990/GRK2379 (IRTG Modern Inverse Problems).
\end{acks}

\bibliographystyle{ACM-Reference-Format}
\bibliography{jackknife}

\end{document}